\PassOptionsToPackage{table,xcdraw}{xcolor}

\documentclass[acmtog]{acmart}

\AtBeginDocument{%
  }

\setcopyright{acmcopyright}
\copyrightyear{2018}
\acmYear{2018}
\acmDOI{XXXXXXX.XXXXXXX}

\acmJournal{TOG}
\acmVolume{37}
\acmNumber{4}
\acmArticle{111}
\acmMonth{8}

\usepackage[linesnumbered,ruled]{algorithm2e} 

\newcommand{\algrule}[1][.2pt]{\par\vskip.5\baselineskip\hrule height #1\par\vskip.5\baselineskip}
\SetAlFnt{\small}
\SetAlCapFnt{\small}
\SetAlCapNameFnt{\small}
\SetAlCapHSkip{0pt}
\IncMargin{-\parindent}

\usepackage[capitalise]{cleveref}
\Crefname{figure}{Figure}{Figures}
\Crefname{table}{Table}{Tables}
\Crefname{section}{Section}{Sections}

\usepackage{graphicx}

\begin{document}

\title{Advancing Front Mapping}

\author{Marco Livesu}
\email{marco.livesu@gmail.com}
\orcid{0000-0002-4688-7060}
\affiliation{%
  \institution{CNR IMATI}
  \city{Genoa}
  \country{Italy}
}
%
%
%
%
%
%
%


\begin{teaserfigure}
\includegraphics[width=\linewidth]{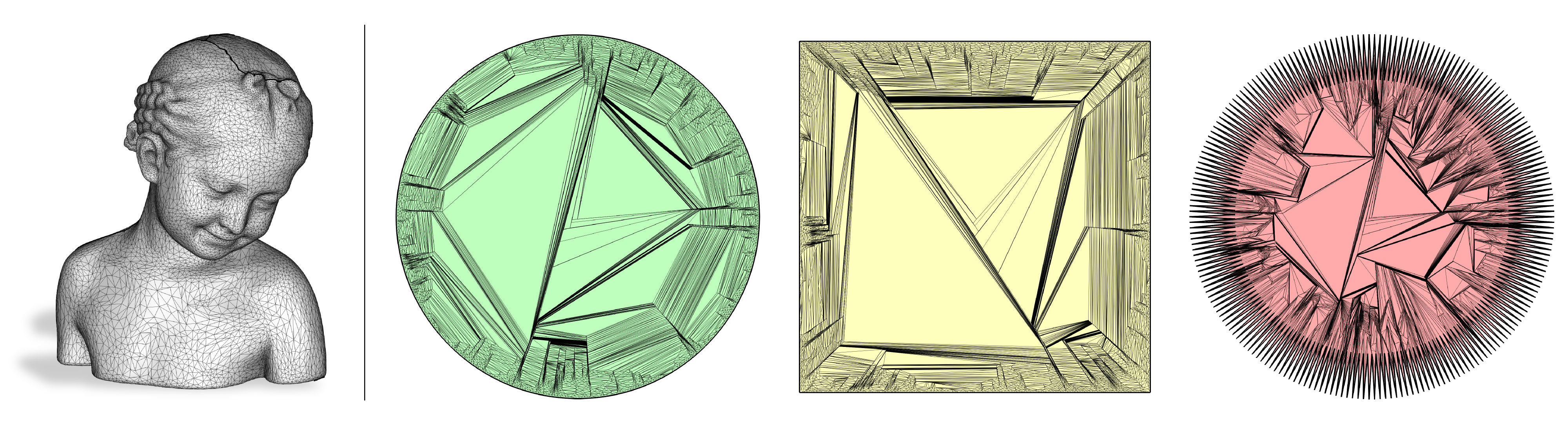}
\caption{Given an input mesh with the topology of a disk (left), the Advancing Front Mapping algorithm allows to flatten it to a strictly convex (green), convex (yellow) or star-shaped (red) domain. All mappings are guaranteed to be free from degenerate or inverted elements, thus being injective.}
\label{fig:teaser}
\end{teaserfigure}

\begin{abstract}

We present Advancing Front Mapping (AFM), a provably robust algorithm for the computation of surface mappings to simple base domains. Given an input mesh and a convex or star-shaped target domain, AFM installs a (possibly refined) version of the input connectivity into the target shape, generating a piece-wise linear mapping between them. The algorithm is inspired by the advancing front meshing paradigm, which is revisited to operate on two embeddings at once, thus becoming a tool for compatible mesh generation. AFM extends the capabilities of existing robust approaches, such as Tutte or Progressive Embedding, by providing the same theoretical guarantees of injectivity and at the same time introducing two key advantages: support for a broader set of target domains (star-shaped polygons) and local mesh refinement, which is used to automatically open the space of solutions in case a valid mapping to the target domain does not exist. AFM relies solely on two topological operators (split and flip), and on the computation of segment intersections, thus permitting to compute provably injective mappings without solving any numerical problem. This makes the algorithm predictable, easy to implement, debug and deploy. We validated the capabilities of AFM extensively, executing more than one billion advancing front moves on 36K mapping tasks, proving that our theoretical guarantees nicely transition to a robust and practical implementation.

\end{abstract}

\maketitle

\section{Introduction}
\label{sec:intro}

Surface mappings are arguably one of the most widely studied topics in computer graphics and are at the core of many fundamental techniques in the field~\cite{floater2005surface,hormann2008mesh,fu2021inversion,naitsat2021inversion}.

In this article we focus on the specific task of mapping a given triangle mesh to a convex or star-shaped domain with fixed boundary. While mappings of this kind are seldom directly useful for downstream applications due to the high geometric distortion they contain (\cref{fig:teaser}), methods for the robust solution of this problem are internally used in many higher level pipelines, which employ this construction to create valid initial maps that are then subsequently improved based on the needs of the application at hand. Various representative examples of this strategy can be found in the literature, spanning from the robust computation of correspondences between shapes~\cite{kraevoy2004cross,schreiner2004inter,shen2019progressive}, to morphing~\cite{praun2001consistent} and the generation of low-distortion planar maps for uv mapping, remeshing, and other applications~\cite{kraevoy2003matchmaker, shen2019progressive,lee1998maps,smith2015bijective,rabinovich2017scalable,liu2018progressive,fargion2022globally,livesu2023towards}.

For a surface map to be practically useful, it is required that no triangle becomes degenerate or inverts its orientation through the map. In other words, mappings must be \emph{injective}. This is a hard property to fulfill. Most of the existing methods achieve a good practical robustness and can perform well in most of the cases, but may unexpectedly fail to produce an injective map. Very few methods can be regarded as being \emph{unconditionally robust}, meaning that they provide strict theoretical guarantees of injectivity (\cref{sec:related}).

\begin{figure}
    \centering
    \includegraphics[width=\columnwidth]{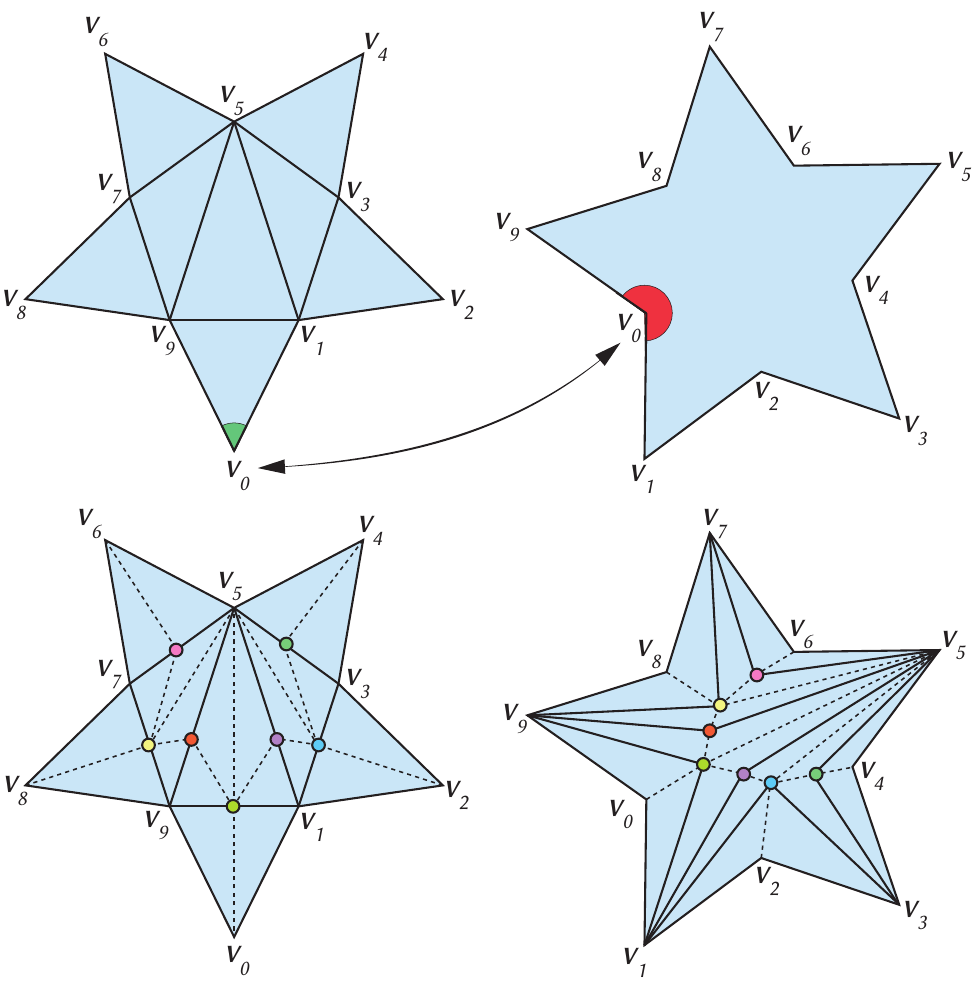}
    \caption{Top: two identical stars up to a shift of the boundary vertices cannot be mapped to one another without introducing additional vertices, because the convexities of the left one map to the concavities of the right one. Bottom: by automatically refining the input mesh our method successfully opens the space of solutions, producing an injective map. Note: due to the presence of tiny triangles in the original output, the inner nodes in the bottom right star have been slightly relocated to make the figure easier to parse.}
    \label{fig:star}
\end{figure}

In this article we introduce a novel methodology for the computation of provably injective mappings to a fixed domain, called \emph{Advancing Front Mapping}. AFM takes inspiration from advancing front mesh generation, which is a meshing strategy that starts from the boundary of a target domain and proceeds inwards, inserting new mesh elements until the whole domain is filled~\cite{george1994advancing}. Our main contribution is an adaptation of this idea to the context of surface map generation, where two fronts are initialized at the boundaries of two alternative domains and are simultaneously advanced while maintaining a one-to-one correspondence between them, reproducing the same topological structure in the two domains. Since a meshing of the input domain is typically known, AFM uses the connectivity of the input mesh as a guidance, trying to install it also in the target domain. This connectivity is a resource and never a limit: in cases a mapping with the given topology does not exist, AFM is able to automatically refine the input mesh, opening the space of solutions and always providing a valid output map (\cref{fig:star}).

Our method is partly inspired by~\cite{marcum1995unstructured} which, similarly to us, uses a background mesh and topological operators to advance the front. However, operating in two domains at once introduces many additional issues which cannot be handled by classical advancing front methods, because they are designed to operate on a single domain only. In the article we discuss and resolve all these issues in a provably robust manner, using a completely novel methodology. Remarkably, we do this by using only a minimal set of geometric and topological constructions. In fact, AFM relies on solely two topological moves to advance the front -- triangle split and edge flip -- and uses simple edge intersections and edge splits to resolve the few deadlock configurations that may arise. As a result, the AFM approach is entirely constructive and completely avoids the use of numerical optimization routines for the computation of the output map. This makes AFM simple to implement, predictable and ultimately easy to debug and deploy.

Compared with existing provably injective approaches such as Tutte~\shortcite{tutte1963draw} and Progressive Embedding~\cite{shen2019progressive}, AFM provides two key advancements: it significantly enlarges the class of supported target domains, permitting mappings to non-convex (star-shaped) polygons, and it also exploits local mesh refinement to open the space of solutions, permitting to generate mappings that would be impossible to compute with prior robust methods due to limits in the mesh connectivity.
In \cref{sec:results} we provide extensive analysis and comparative evaluation with prior art, also showcasing both the theoretical and practical robustness of our approach.




\section{Related Works}
\label{sec:related}

Our contribution is focused on 
a very specific instance of the (more general) surface mapping problem. In the remainder of this section we discuss only methods that can compute mappings to domains with a fixed boundary. We point the reader to~\cite{hormann2008mesh} for a broader perspective on general surface mappings and their related applications.

\subsection{Provably Robust Methods}
\label{sec:related_robust}
Only a few methods in the literature offer strict theoretical guarantees of correctness, meaning that injectivity can be formally proven (e.g. with a theorem). Note that theoretical guarantees do not necessarily lead to practical robustness. In fact, when implemented with limited precision floating points, robust methods can still fail to produce a correct result, introducing vanishing or inverted elements in the map~\cite{shen2019progressive}. Conversely, when implemented using exact numerical models (e.g., rational numbers), methods in this class are guaranteed to always produce an injective map.

\paragraph{Tutte-like embeddings} Solutions to the problem of mapping a surface to a strictly convex domain have been known since 1963, when Tutte showed that positioning internal vertices at the barycenter of their neighbors yields a valid embedding~\cite{tutte1963draw}. This approach was brought to the computer graphics community and extended by Floater, who showed that the theorem still holds when the boundary is not strictly convex and when interior vertices are positioned at a general convex combination of their neighbors~\cite{floater1997parametrization,floater2003one,floater2003mean}. Additional flexibility was provided in~\cite{gortler2006discrete}, where the authors show that under some topological restrictions on the mesh connectivity a "Tutte-like" embedding could also be proved for a subset of meshes containing multiple boundaries, possibly non convex. While meshes included in this category are not even required to be star-shaped, overall these topological requirements are tight and the set of non convex domains supported by this theoretical framework covers just a restricted set of non convex shapes. A similar extension (though with a completely different formulation) was recently discussed in~\cite{kovalsky2020non}, where the authors use the notion of \emph{cone condition} to express additional positional constraints for inner vertices that are incident to non convex boundary vertices. As shown in \cref{fig:star} mappings to non convex domains may not be possible if the mesh connectivity is fixed. Neither~\cite{gortler2006discrete} nor~\cite{kovalsky2020non} permit to change the input mesh, meaning that if a solution does not exist these methods will output a non injective map. Establishing \emph{a priori} whether the solution space is vanishing or not and devising a restricted set of topological changes that ensure the existence of a solution is a complex matter, thus making these methods intrinsically trial-and-error.

\paragraph{Intermediate domains.} 
A successful strategy to enlarge the class of domains onto which a mesh can be mapped is the composition of multiple maps. Methods in this category internally use the Tutte embedding, generating mappings to strictly convex domains that are then combined to obtain more general mappings to non-convex planar polygons~\cite{weber_zorin14}, manifolds of arbitrary genus~\cite{livesu2020scalable,li2008globally,garner2005topology}, or grid spaces for quad remeshing~\cite{livesu2023towards}. These methods do not introduce novel methodologies for mappings to a fixed boundary, therefore more than competitors they can be regarded as candidate users of our construction, which could successfully substitute Tutte for the generation of the intermediate mappings, retaining the same theoretical guarantees of correctness and at the same time enlarging the class on intermediate domains that can be used.

\paragraph{Progressive Embedding (PE)} In~\shortcite{shen2019progressive} Shen and colleagues observed that the Tutte embedding is prone to failures when implemented in limited precision and proposed an alternative approach, called Progressive Embedding, which offers the same theoretical guarantees of correctness but also higher practical robustness when implemented with floating points. Inspired by the concept of Progressive Meshes~\cite{hoppe1996progressive}, this method operates by first removing all flipped elements from an input non injective map through a sequence of edge collapses. Then, it reintroduces the previously removed vertices one at a time, through a sequence of vertex splits that inverts the sequence used in the simplification phase. Newly inserted vertices are carefully positioned in the domain, ensuring that no incident triangle inverts its orientation or vanishes, thus preserving the injectivity of the map. Despite not mentioned by the authors and not supported by their reference implementation, we believe that PE could in principle produce mappings to star-shaped domains. In fact, removing all internal vertices but one in the simplification phase and positioning such vertex inside the kernel of the target star-shaped domain, all vertices could be reintroduced with their vertex split procedure without violating  injectivity. However, since this method does not allow to change the input mesh connectivity, cases where a mapping is impossible without mesh refinement (e.g., \cref{fig:star}) would still lead to failures. PE heavily exploits Newton-like optimization to regularize the embedding and produce an injective map in floating points, a task at which it remains unbeaten (\cref{sec:results}). However, this makes it three orders of magnitude slower than the approach we propose which, conversely, completely avoids the resolution of numerical problems.

\paragraph{Foliations} Campen and colleagues introduced the concept of simplicial foliations~\shortcite{campen2016bijective}, a method to map both 2D and 3D simplicial meshes to simple base domains. Also this method is provably robust, but it only supports mappings to a squares and circles, also not permitting to setup explicit per vertex boundary conditions. Foliations yield maps that are non linear inside each triangle, which can be transformed into piece-wise linear maps only at the expense of massive mesh refinement, increasing mesh size even by orders of magnitude. 

\paragraph*{Compatible Triangulations} Methods that fill two empty domains with a compatible triangulation are similar in spirit to our method, in the sense that they share the same mesh generation perspective on the computation of mappings. In~\shortcite{Liv20b} Livesu showed how a simple triangulation algorithm (earcut) may be exploited to construct a provably injective mapping between a general polygon and a convex target domain. A generalization of the base algorithm that also supports mappings to star-shaped domains is proposed in~\cite{livesu2020mapping}. Methods such as~\cite{surazhsky2004high,gupta1997constructing} are even more powerful, permitting to construct compatible triangulations between any two simple concave polygons. All these methods can only operate in a 2D-to-2D setting. The case in which the source mesh is a surface embedded in a higher dimensional space is not supported. The concept of compatible triangulations was recently lifted to the surface case~\cite{takayama2022compatible}, but this latter method requires to be initialized with a previously existing map, hence it does not solve the problem we are interested in.\\

\begin{figure*}[h]
	\includegraphics[width=\linewidth]{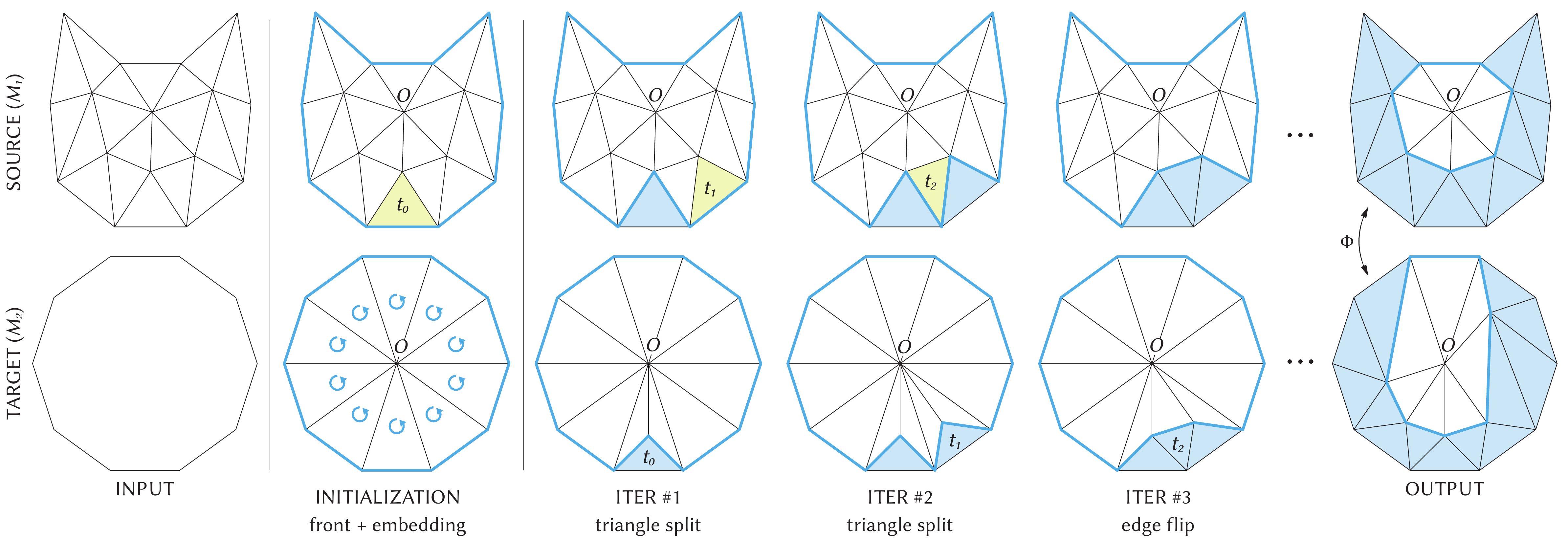}
	\caption{Pipeline of AFM: we start from a source triangle mesh and a target convex or star-shaped domain (left). In the initialization step (second column) boundary edges are marked as front (thick blue lines) and a starting embedding is created by forming a triangle between each front edge and the front origin ($O$), selected as an inner mesh vertex. Due to convexity, all triangles share a globally coherent orientation. Then, AFM iteratively reproduces the connectivity of the source mesh in the target domain, advancing both fronts while maintaining a one-to-one correspondence between them. Source triangles having only one edge on the front ($t_0$ and $t_1$) are inserted by splitting the triangle formed by the image of such edge and $O$ (iterations \#1 and \#2). Source triangles having two edges on the front ($t_2$) are inserted by flipping the edge connecting $O$ with the front vertex shared by the images of such edges (iteration \#3). The algorithm stops when all source triangles have been inserted in the target domain, yielding a one-to-one mapping $\Phi$ between the two meshes.}
	\label{fig:pipeline}
\end{figure*}

\subsection{Numerical Methods} A large number of methods can compute planar mappings to an arbitrary domain with fixed boundary by solving for the 2D coordinates of inner mesh vertices, subject to hard positional constraints for boundary vertices. Solving this problem naively can easily introduce flipped elements in the mapped mesh. On the other hand, explicitly imposing injectivity constraints yields a non-linear non-convex optimization problem that cannot be solved directly~\cite{weber_zorin14}. Various techniques exist to relax this formulation and make it numerically tractable, using barrier energies, convexification of the feasible space through map projection or simplex assembly. We point the reader to~\cite{fu2021inversion,naitsat2021inversion} for a detailed description of these techniques. The most recent literature in this field employ flip-preventing energies that promote low geometric distortion and at the same time avoid the generation of flipped elements, exhibiting a remarkable ability to produce injective maps even in intricate cases~\cite{du2020lifting,garanzha2021foldover,poya2023geometric}. However, operating in the whole feasible space of solutions with explicit injectivity constraints remains unfeasible, hence numerical approaches cannot provide strict guarantees of correctness, making them less reliable than methods reviewed in \cref{sec:related_robust} and than the advancing front strategy that we propose. These methods also do not natively support on demand mesh refinement, possibly triggering unpredictable behavior when they are required to solve a mapping problem for which a solution does not exist.

\subsection{Hybrid Methods} Methods in this class take the best of both worlds, combining the unconditional robustness of provably injective approaches and the low geometric distortion of numerical approaches. The main idea -- shared by all existing techniques -- is to use the Tutte embedding to initialize a provably injective map and then reduce geometric distortion by carefully minimizing an energy functional, using line search and rollback operators to make sure that no vertex move breaks the injectivity of the map. This paradigm was pioneered in~\cite{smith2015bijective} and subsequently adopted by numerous authors~\cite{rabinovich2017scalable,liu2018progressive,fargion2022globally,jiang2017simplicial}. Similarly to robust methods that extend Tutte by using maps to intermediate domains, also these methods can be regarded as potential users of the technique we present. In fact, AMF could substitute Tutte for the generation of the initial injective map, likely providing a better starting solution with lower distortion, thanks to the additional degrees of freedom that permit to robustly map the input shape also to non convex domains.


\section{Method}
\label{sec:method}
AFM takes in input a triangle mesh with disk-like topology, $M_1$, and an injective mapping of its boundary vertices onto the boundary of a target domain $\Omega$. Such domain is required to be convex or star-shaped, meaning that it contains a non empty kernel from which all its boundary vertices are directly visible. The algorithm outputs two meshes: a possibly refined version of $M_1$ and a mesh $M_2$ that triangulates $\Omega$ and has the same connectivity of $M_1$.

Since the two meshes contain the same number of vertices, connected to form the same triangles, there exists a natural piece-wise linear mapping between them
$$\Phi : M_1 \leftrightarrow M_2.$$
Function $\Phi$ is constructed through a direct (per vertex) correspondence and is linearly extended inside mesh triangles via barycentric interpolation. Similarly to~\cite{tutte1963draw,shen2019progressive}, AFM is guaranteed to always produce a piece-wise linear injective map, meaning that neither $M_1$ nor $M_2$ contain zero-area or inverted elements, hence $\Phi$ is one-to-one.\\

At a high level the algorithm is rather simple: after having created a polar mesh $M_2$ that covers the target domain $\Omega$, the boundary edges of $M_1,M_2$ are set to form two initial fronts with simple topology (\cref{fig:pipeline}, Initialization). These two fronts are then coherently pushed inwards, maintaining a one-to-one correspondence throughout the whole execution. Each advancing front move amounts to conquering an unvisited triangle in $M_1$, inserting an equivalent triangle in $M_2$. Advancing moves in the target domain are always executed by either splitting an existing triangle in $M_2$ or by flipping one of its edges (\cref{fig:pipeline}, Iterations). Both moves are guaranteed to not create degenerate or inverted elements if simple local geometric requirements are fulfilled, thus ensuring the injectivity of the map.
In case of geometric deadlock configurations that prevent the front to advance any further, vertex relocation or finite local mesh refinement (applied to both meshes) is used to resolve the lock and proceed with the computation. Refinement is used parsimoniously and only marginally impacts the input mesh size (\cref{sec:results}). 
The algorithm stops when all triangles of $M_1$ have been successfully inserted into mesh $M_2$, yielding two meshes with identical connectivity and different embedding.  An algorithmic description of AFM is reported in \cref{alg:AFM}. In the remainder of this section we discuss details of each step, including initialization, front advancement, and handling of the deadlock configurations that may arise.

%
%


\begin{algorithm}
	\SetKwData{Left}{left}
	\SetKwData{Up}{up}
	\SetKwFunction{FindCompress}{FindCompress}
	\SetKwInOut{Input}{inputs}
	\SetKwInOut{Output}{outputs}
	\Input{(1) a triangle mesh $M_1$ with the topology of a disk; (2) a one-to-one correspondence between the boundary vertices of $M_1$ and the boundary of a convex or star-shaped domain $\Omega$.}
	\BlankLine	
	\Output{(1) a possibly refined version of $M1$; (2) a new mesh $M_2$ which has the same connectivity of $M_1$ and that realizes an injective mapping of $M_1$ onto the target domain $\Omega$.}
	\algrule[1pt]
	\SetAlgoLined
	refine $M_1$; \hfill (\cref{sssec:refinement})\\
	mark all triangles in $M_1$ as unvisited;\\
	select the front origin $O$ and initialize $M_2$; \hfill (\cref{sssec:background_mesh})\\
	initialize front as the set of boundary edges in $M_1$;\\
	push all front edges into a queue $Q$;\\
	\vspace{0.3em}
	\While{$Q$ is not empty}
	{
		\vspace{0.3em}
		pop edge $e$ from $Q$;\\
		select the triangle $t \in M_1$ inside the front and incident to $e$;\\
		
		\vspace{0.3em}
		\If{$t$ is already marked as visited}
		{
			\vspace{0.3em}
			\textbf{continue;}\\
		}
		\vspace{0.3em}
		let $E$ be the front edges incident to $t$\\
		\vspace{0.3em}
		\If{$\vert E \vert$ = 1}
		{
			\vspace{0.3em}
			let $v$ be the vertex of $t$ opposite to $e$;\\
			\If{$v$ is on the front or is the front origin $O$}
			{
				\vspace{0.3em}
				\textbf{continue;}\\
			}
			advance front in $M_2$ by triangle split; \hfill (\cref{sssec:one_front_edge})\\
		}
		\ElseIf{$\vert E \vert$ = 2}
		{
			\vspace{0.3em}
			\If{edges in $E$ form a concave angle}
			{
				\vspace{0.3em}
				locally convexify front;\hfill (\cref{sssec:convexification})
			}
			\vspace{0.3em}
			let $t_E$ be the triangle denoted by edges in $E$;\\
			\If{$t_E$ contains the front origin $O$}
			{
				\vspace{0.3em}
				locally concavify front; \hfill (\cref{sssec:concavification})
			}
			\vspace{0.3em}
			advance front in $M_2$ by edge flip; \hfill (\cref{sssec:two_front_edges})\\
		}
		\vspace{0.3em}
		mark $t$ as visited;\\
		push new front edges in $Q$;\\
	}
	\caption{Advancing Front Mapping (AFM)}
	\label{alg:AFM}
\end{algorithm}

\subsection{Initialization}
\label{sec:init}
In this phase the input mesh $M_1$ is first refined to avoid predictable degenerate configurations. Then,  mesh $M_2$ is initialized as a polar mesh that entirely covers the target domain $\Omega$. After these operations have been executed, the algorithm is ready for the iterative part and the boundaries of meshes $M_1$ and $M_2$ form the two fronts to be advanced.

\subsubsection{Refinement}
\label{sssec:refinement}
From a topological perspective, advancing a front with simple topology starting from the boundary and proceeding inwards until the whole mesh is conquered corresponds to computing a shelling sequence of $M_1$. For topological disks, a shelling sequence is always guaranteed to exist if the mesh graph  is 2-connected~\cite{campen2016bijective}(\S4.4). This property can be easily enforced by splitting all internal edges that connect pairs of boundary vertices in the input mesh. This edge splitting strategy also provides additional benefits to our algorithm, because:
\begin{itemize}
    \item it ensures that the mesh has at least one internal vertex, which is a  necessary condition for the initialization of mesh $M_2$;
    \item it ensures that no triangle has all its three vertices on the boundary, which may become flipped or degenerate when mapping to concave or non strictly convex domains such as a square (due to concavity or co-linearity of the boundary edges).
\end{itemize}

\subsubsection{Polar Mesh}
\label{sssec:background_mesh}
For the final mapping to be injective, AFM needs to insert the connectivity of $M_1$ onto the mesh $M_2$ by always operating onto a \emph{valid} mesh. In our setting a mesh is valid if all triangles share a globally coherent orientation and are not degenerate. Constructing such a mesh is trivial if $\Omega$ is convex or star-shaped. In fact, in case of convexity $M_2$ can be initialized by taking any point strictly inside $\Omega$ and forming a triangle between such point and each boundary edge.  Similarly, if the domain is star-shaped a valid mesh $M_2$ can be created with the same procedure, with the additional restriction that the inner point must be located inside the kernel of the domain, that is, any point inside $\Omega$ from which all boundary vertices of $M_2$ are directly visible~\cite{sorgente2022polyhedron}. A visual example of the so generated mesh can be seen in \cref{fig:pipeline}, at the bottom of the second column. Since this mesh contains only one internal vertex that is incident to all boundary triangles, we call such vertex the \emph{front origin}, and denote it with the letter $O$. 

We recall that $M_1$ and $M_2$ will eventually contain the same connectivity, therefore an interior vertex in mesh $M_1$ must also be selected and put in correspondence with $O$. Thanks to the preliminary refinement step, an interior vertex in $M_1$ is guaranteed to exist. From a theoretical perspective, any interior vertex is equally good. In practice, using a point that is furthest from the boundary promotes a uniform growth of the front, providing higher numerical stability. We select such vertex by simply running the Dijkstra shortest path algorithm with multiple sources, selecting all boundary vertices as source nodes and flooding the whole mesh until the furthest vertex is found.

\subsection{Advancing Moves}
\label{sec:advancing_moves}
This is the core of the algorithm. All front edges of the input mesh $M_1$ are inserted into a queue and then processed one at a time until the queue becomes empty. Given a front edge $e$, AFM aims to insert into the target domain the image of the triangle $t$ that is incident to $e$ from the inner side of the front. Two cases are possible: $t$ may have either one or two of its edges exposed on the front (including $e$). Note that the case where all the three edges of $t$ are on the front cannot happen, because the front has simple topology and there is at least one vertex inside the front (the origin $O$).

\begin{figure}
	\includegraphics[width=\columnwidth]{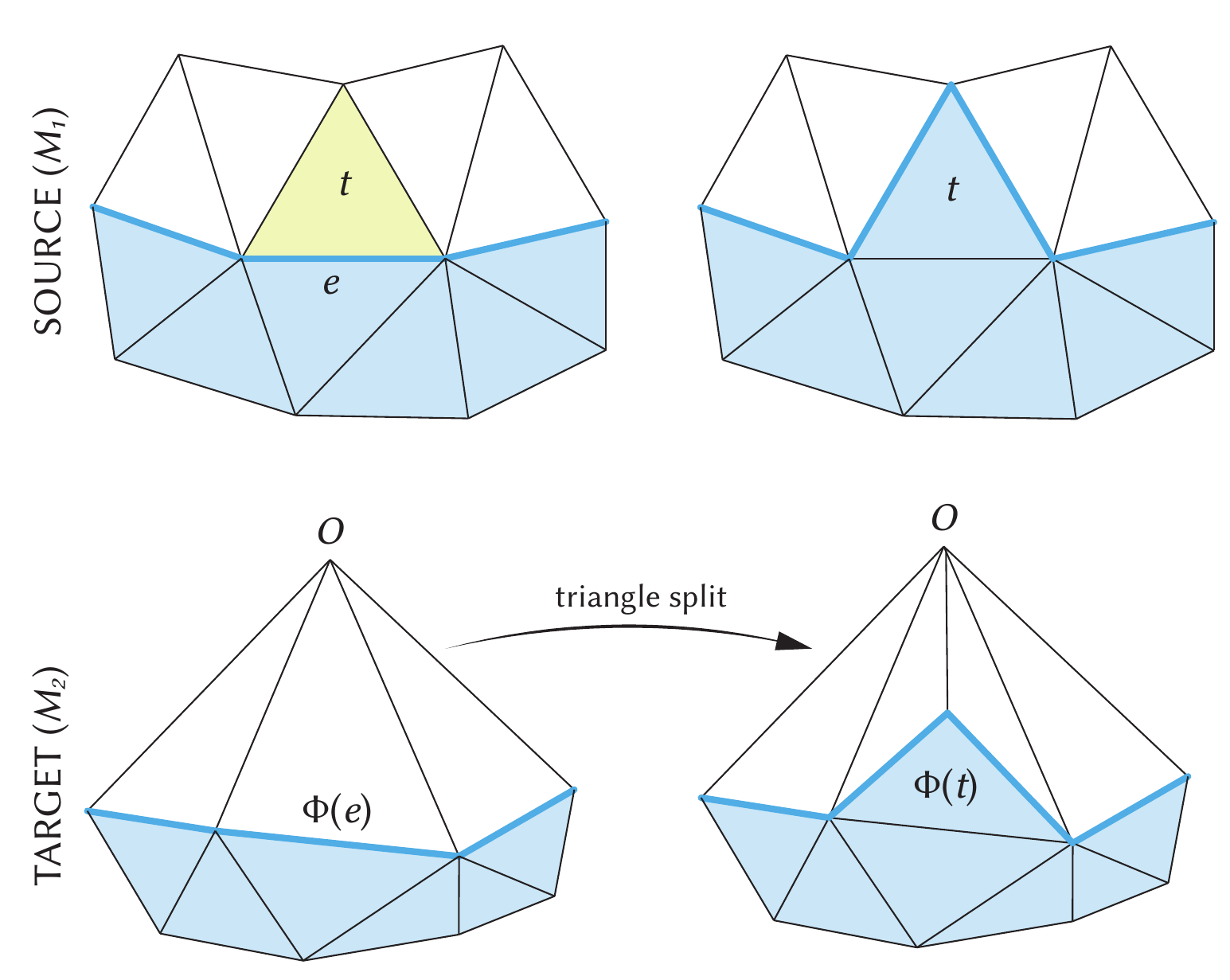}
	\caption{Advancing the front to reproduce a triangle $t$ having a single front edge $e$ amounts to: (i) locating the triangle formed by the image of $e$, $\Phi(e)$, and the front origin $O$; (ii) splitting it into three sub-triangles. The triangle formed by the split point and by $\Phi(e)$ is $\Phi(t)$.}
	\label{fig:triangle_split}
\end{figure}

\subsubsection{One front edge}
\label{sssec:one_front_edge}
Let $e$ be the only edge of $t$ exposed on the front and $v$ the vertex of $t$ opposite to it. If vertex $v$ is inside the front and has not been inserted into mesh $M_2$ yet, we insert the image of $t$ into mesh $M_2$ by splitting the triangle formed by the image of $e$ and the front origin, $O$. We do this by using as a split point the image of vertex $v$, $\Phi(v)$ (\cref{fig:triangle_split}). This operation produces three sub-triangles, one of which is formed by $\Phi(e)$ and $\Phi(v)$, hence is the image of triangle $t$. The front in both meshes is then advanced, moving from $e$ to the other two edges of triangle $t$, which is now located outside of the current front (see iterations $\#1$ and $\#2$ in \cref{fig:pipeline}). Note that, since triangle $t$ has been inserted into $M_2$ by means of a triangle split, the newly generated triangles are guaranteed to preserve the orientation of their father, thus not breaking local injectivity. For the specific location of point $\Phi(v)$ in $M_2$,  any point strictly inside the triangle formed by $\Phi(e)$ and $O$ is guaranteed to produce a valid mesh. In our implementation we position $\Phi(v)$ as
$$\Phi(v) = \frac{99\:\Phi(v_0) + 99\:\Phi(v_1) + 2O}{200},$$
where $v_0$ and $v_1$ are the vertices of edge $e$. This choice avoids the front to rapidly approach the origin $O$, leaving more room for the insertion of the subsequent mesh elements, thus improving numerical stability.

\begin{figure}
	\includegraphics[width=\columnwidth]{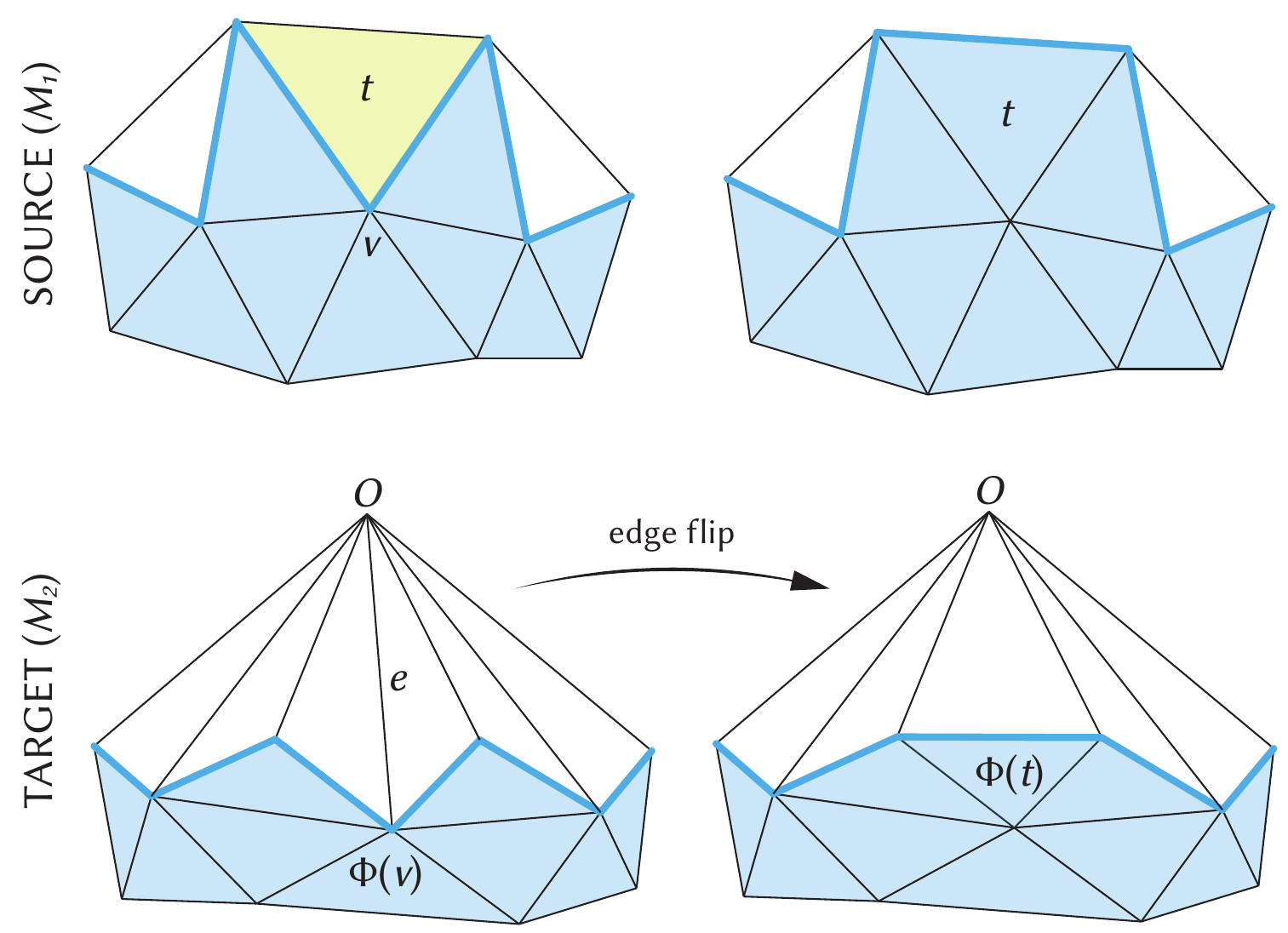}
	\caption{Advancing the front to reproduce a triangle $t$ having two front edges meeting at a shared vertex $v$ amounts to: (i) locating the edge $e$ connecting the image of $v$, $\Phi(v)$, with the front origin $O$; (ii) flipping $e$. This operation can be performed only if the quad surrounding $e$ (i.e., its \emph{link}) is strictly convex.}
	\label{fig:edge_flip}
\end{figure}

\subsubsection{Two front edges}
\label{sssec:two_front_edges}
Let $e_1,e_2$ be the two edges of $t$ exposed on the front and let $v$ be the front vertex in between them. In the target domain edges $\Phi(e_1)$ and $\Phi(e_2)$ form two triangles with the front origin $O$ and are adjacent to one another along the edge connecting $\Phi(v)$ with $O$. The insertion of triangle $t$ inside mesh $M_2$ can therefore be performed by simply flipping such edge (\cref{fig:edge_flip}). The front in both meshes is then advanced, moving from $e_1,e_2$ to the edge of $t$ opposite to them (see iteration $\#3$ in \cref{fig:pipeline}). Note that, differently from the triangle split, the edge flip operation cannot always be performed.
Deadlock cases that arise when no further flip move is possible are discussed in the next section.

\begin{figure}
	\includegraphics[width=\columnwidth]{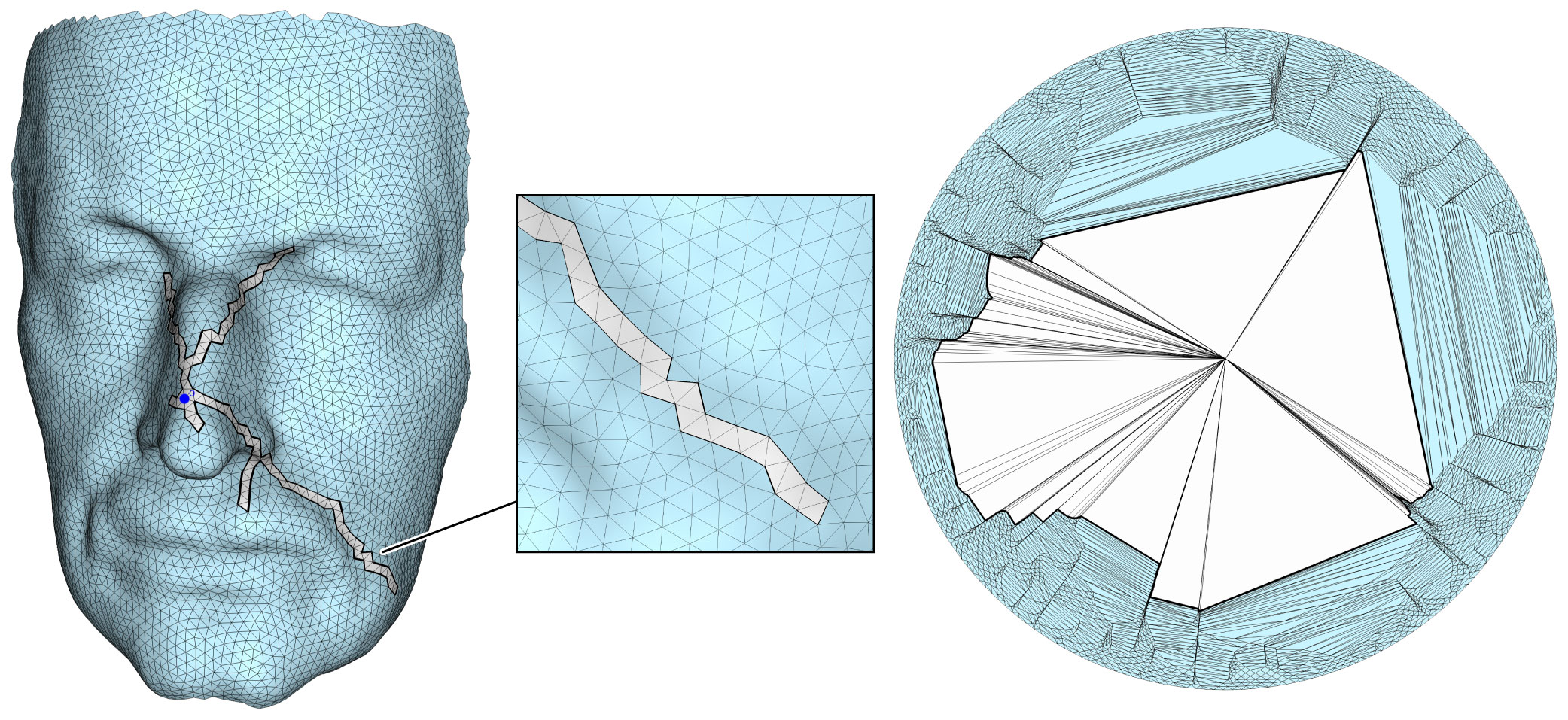}
	\caption{Naively applying triangle splits and edge flips to advance the front does not always lead to convergence. In the typical case, the front conquers all the mesh vertices, leaving a network of triangle strips (closeup) that cannot be inserted in the target domain because their associated edge flip operations are concave, hence they would introduce flipped elements. The local strategies described in \cref{sec:corner_cases} ensure that any edge flip can be performed without breaking the injectivity of the map, thus always ensuring both validity and convergence.}
	\label{fig:max_no_refinement}
\end{figure}

\begin{figure*}[h]
	\includegraphics[width=\linewidth]{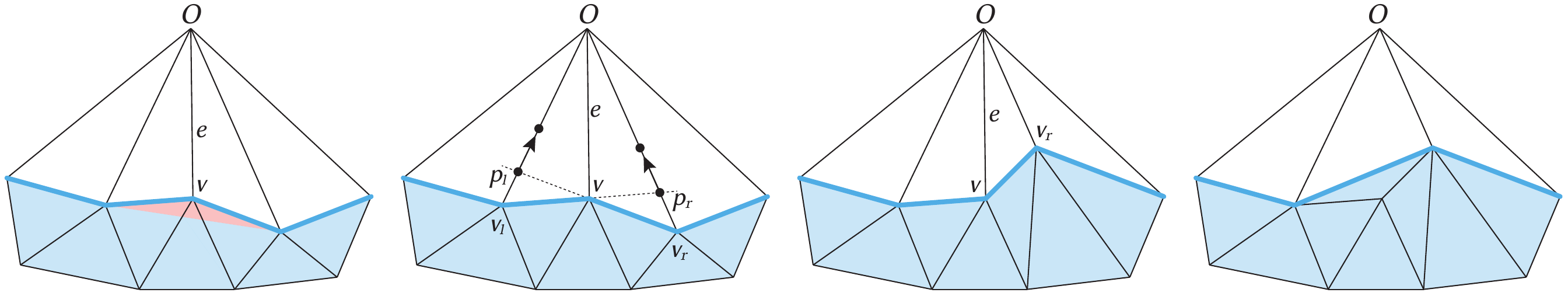}
	\caption{Left: flipping edge $e$ introduces an inverted element (red) because the front is locally concave at vertex $v$. Middle left: we compute points $p_l,p_r$ by intersecting the prolongation of edges $v,v_l$ and $v,v_r$ with edges $v_l,O$ and $v_r,O$. Positioning either $v_l$ or $v_r$ at such intersection points makes the front locally flat. Pushing any of them a bit further towards the origin (black arrows) makes the front locally convex at $v$. Middle right: since only one vertex motion is necessary, we break ties by moving the vertex that remains furthest from $O$ (in this case $v_r$), leaving more room for the next advancing moves. For the same reason, in our implementation we only minimally lift vertex $v_r$, computing its final position as $0.99p_r+0.01O$. Right: after vertex relocation the front can be advanced by flipping edge $e$.}
	\label{fig:convexification}
\end{figure*}

\begin{figure*}[h]
	\includegraphics[width=\linewidth]{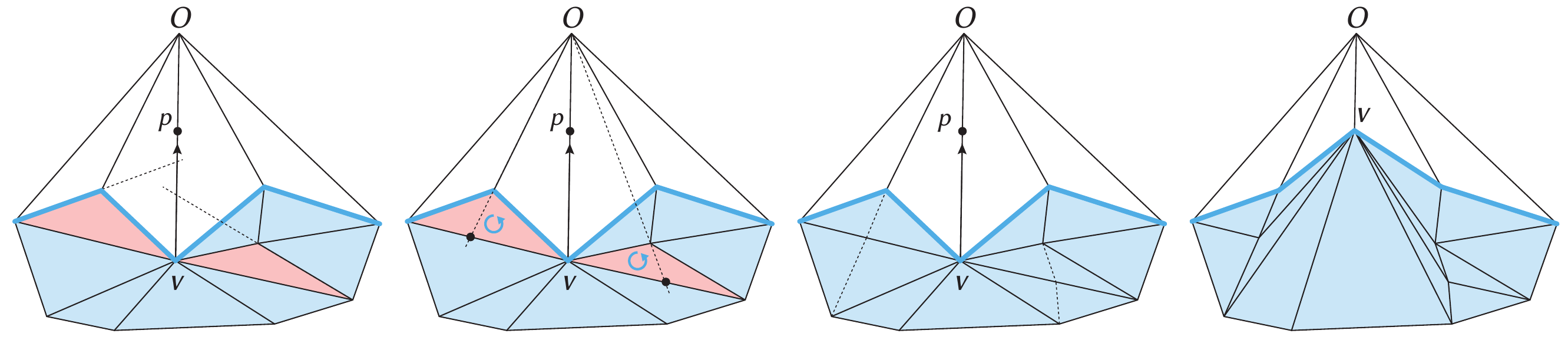}
	\caption{Left: pushing vertex $v$ towards point $p$ flips the orientation of some triangles (in red) because, for each of them, $v$ and $p$ lie at opposite sides of the supporting line of the edges opposite to $v$ (dashed lines). Middle left: split points to fix this issue can be computed by intersecting, for each red triangle, the edge whose positive half-space contains the front origin $O$ with the line passing through $O$ and the triangle vertex opposite to such edge. Middle right: splitting edges at the so computed intersections yields a new mesh where all triangles remain valid when pushing $v$ towards $O$ (right).}
	\label{fig:convexification_refinement}
\end{figure*}

\subsection{Deadlocks}
\label{sec:corner_cases}
An edge can be flipped without introducing inverted elements only if the four-sided polygon formed by the two triangles incident to it is strictly convex. This may create deadlock configurations where the front can no longer be advanced, because all missing triangles have their three vertices exposed on the front and no edge flip move is possible. Situations of this kind typically arise towards the end of the execution, when all input mesh vertices have been inserted in the target domain and a sequence of edge flips is required to \emph{close} the front (\cref{fig:max_no_refinement}). 

Concavities that prevent the execution of a flip move may arise at both endpoints of the edge to be flipped. In AFM flipped edges always connect a front vertex with the origin $O$. In the following two subsections we describe the two procedures that we use to unlock illegal edge flips in these two cases. Thanks to these procedures, \emph{any} edge flip can be executed to reproduce a mesh triangle of $M_1$ in $M_2$, thus guaranteeing convergence.


%

\subsubsection{Convexification}
\label{sssec:convexification}
When the front is locally concave at a front vertex $v$, we unlock the flip of the edge connecting $v$ and $O$ via vertex relocation, pushing one of the front vertices adjacent to $v$ towards $O$, making the front locally convex at $v$.
The way we determine which of the two front neighbors of $v$ should be moved and by how much is described in \cref{fig:convexification}. 
Once the selected point has been repositioned to the desired location, the edge can be flipped and the front advanced as described in \cref{sssec:two_front_edges}.

\begin{figure*}
	\includegraphics[width=\linewidth]{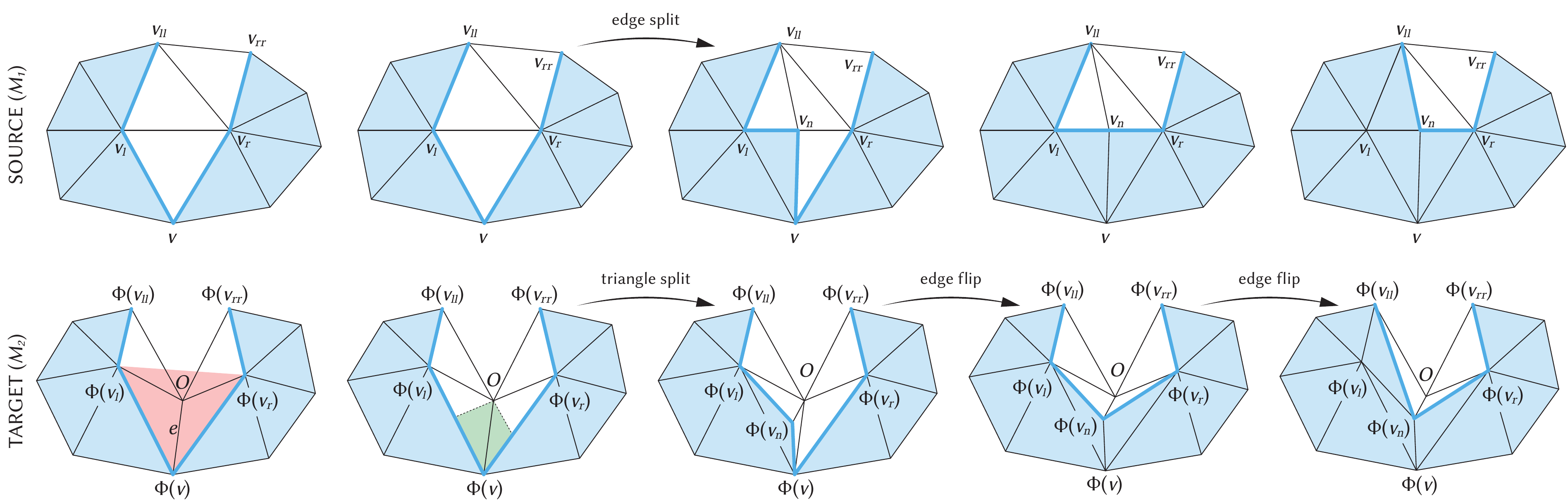}
	\caption{Left: flipping edge $e$ introduces an inverted element (red) because the triangle $\Phi(v_l,v,v_r)$ contains the front origin $O$. We resolve this issue by inserting a new vertex $v_n$ in both meshes, locally editing the front to avoid advancing it beyond the origin $O$. In the input domain vertex $v_n$ splits the edge $v_l,v_r$ at its midpoint. In the target domain, its image $\Phi(v_n)$ is positioned so as to create valid edge flip moves to insert triangles $\Phi(v_n,v,v_r)$ and $\Phi(v_l,v_n,v_{ll})$. The feasible region is highlighted in the second column (green area) and is computed by intersecting the prolongations of edges $\Phi(v_r),O$ and $\Phi(v_{ll}),O$ with the front. The advancing moves are shown in the last three columns. Note that in case $\Phi(v_n)$ is not inserted in the green region, one or both the edge flip moves in the last two columns would in turn suffer from the same issue shown in the first column, thus triggering infinite refinement.}
	\label{fig:concavification}
\end{figure*}

It should be noted that the convexification strategy described in \cref{fig:convexification} is not completely safe, because it only ensures that the two triangles generated by the edge flip are not inverted, completely overlooking the effect that this operation has on all the other triangles that are incident to a relocated vertex $v$. As shown in the left part of \cref{fig:convexification_refinement}, previously existing triangles (in red) may flip their orientation due to vertex relocation. We resolve this issue by splitting such triangles as described in the same figure, ensuring that, for each triangle $t$ incident to $v$, the positive half-space of the edge $e$ opposite to $v$ in $t$ contains the front origin $O$. Since vertices are always relocated at a point along an edge incident to $O$, this ensures that any possible relocation does not introduce inverted elements in $M_2$. Note that, differently from the triangle splits in \cref{sssec:one_front_edge}, the refined triangles are images of triangles existing in the input mesh. Therefore, these splits are also applied to triangles in $M_1$, maintaining a full topological compatibility with mesh $M_2$.

To avoid excessive mesh growth, convexification should be used parsimoniously. AFM is designed to promote the fulfillment of the geometric criteria that ensure the applicability of flip operators and only marginally uses convexification. In our large scale benchmark we performed almost half a billion edge flips. Only in the 15.7\% of the cases convexification was necessary and only in the 3.7\% of these cases local refinement was used to unlock a flip (\cref{sec:results}).

Another important aspect to consider is convergence. Is this refinement always finite? The answer is yes. Refinement provably converges in a finite number of steps because the number of triangles incident to $v$ is finite, yet because the refinement does not change the vertex valence. In particular, it can be shown that if offending triangles are processed starting from the front and proceeding inwards along the topological triangle sorting around $v$, then the maximum number of splits corresponds with the valence of vertex $v$, which is typically a small number (on average 6 on regular meshes).

\subsubsection{Concavification}
\label{sssec:concavification}
When the concavity that prevents a flip is located at the front origin $O$, we unlock the flip operation by adding one degree of freedom along the front, splitting the triangle that would flip its orientation (left of \cref{fig:concavification}, in red) into two sub-triangles that do not contain $O$. We call this procedure \emph{concavification}, because a portion of the front that is already convex becomes locally concave to accommodate the flip move that we need to perform. Concavification is more critical than the convexification strategy discussed before because, depending on the positioning of the newly inserted vertex, it may generate infinite refinement. The key to successfully resolve the issue and ensure convergence consists in positioning the new vertex not only in function of the current advancing move, but also considering the next two, finding a location that ensures that a full sequence of three advancing moves can be performed without any further convexification or concavifications (last three columns of \cref{fig:concavification}). In terms of impact on the output mesh size, concavification is even more rare to occur than convexification. In our large scale benchmark this procedure was applied in only the 0.29\% of the cases.

A tempting (apparently simpler) alternative to concavification consists in relocating vertex $O$ outside of the red triangle in~\cref{fig:concavification}. While in some cases this would work without requiring extra refinement, it should be noted that not only $O$ but also other portions of the front may sneak inside that triangle, possibly requiring a large number of relocations, each of which may trigger its own refinement (\cref{fig:convexification_refinement}). Considering the marginal number of its occurrences, we opted for concavification because it is compact to code and it always has the same complexity, avoiding the necessity to handle challenging corner cases in the code.

\begin{table*}[h]
	\resizebox{.9\textwidth}{!}{%
		\begin{tabular}{l|ccc|ccccc|c|cc|c}
			\rowcolor[HTML]{FFFFFF} 
			\textbf{Domain} &
			\textbf{\#Models} &
			\textbf{\#Conv.} &
			\textbf{\#Timeout} &
			\textbf{\begin{tabular}[c]{@{}c@{}}\#Adv.\\ Moves\end{tabular}} &
			\textbf{\begin{tabular}[c]{@{}c@{}}\#Triangle\\ Splits\end{tabular}} &
			\textbf{\begin{tabular}[c]{@{}c@{}}\#Edge\\ Flips\end{tabular}} &
			\textbf{\#Convex.} &
			\textbf{\#Concav.} &
			\textbf{\begin{tabular}[c]{@{}c@{}}Mesh\\ growth\end{tabular}} &
			\textbf{\begin{tabular}[c]{@{}c@{}}\#Flips\\ (rational)\end{tabular}} &
			\textbf{\begin{tabular}[c]{@{}c@{}}\#Flips\\ (double)\end{tabular}} &
			\textbf{\begin{tabular}[c]{@{}c@{}}Average\\ runtime\end{tabular}} \\ \hline
			\rowcolor[HTML]{FFFFFF} 
			Circle &
			11942 &
			11679 &
			\begin{tabular}[c]{@{}c@{}	}263\\96.5\% done\end{tabular} &
			314.6M &
			\begin{tabular}[c]{@{}c@{}}152.7M\\48.5\%\end{tabular} &
			\begin{tabular}[c]{@{}c@{}}161.9M\\51.5\%\end{tabular} &
			\begin{tabular}[c]{@{}c@{}}15.1M\\9.3\%\end{tabular} &
			\begin{tabular}[c]{@{}c@{}}918.6K\\0.56\%\end{tabular} &
			\begin{tabular}[c]{@{}c@{}}avg 4.91\%\\max 42.6\%\end{tabular} &
			0 &
			\begin{tabular}[c]{@{}c@{}}885.5K\\ 3166 models\\ 27.1\%\end{tabular} &
			0.77s \\ \hline
			\rowcolor[HTML]{EFEFEF} 
			Square &
			11942 &
			11645 &
			\begin{tabular}[c]{@{}c@{}}297\\ 96.2\% done\end{tabular} &
			312.3M &
			\begin{tabular}[c]{@{}c@{}}151.6M\\48.5\%\end{tabular} &
			\begin{tabular}[c]{@{}c@{}}160.7M\\51.5\%\end{tabular} &
			\begin{tabular}[c]{@{}c@{}}15.7M\\9.8\%\end{tabular} &
			\begin{tabular}[c]{@{}c@{}}910.1K\\0.57\%\end{tabular} &
			\begin{tabular}[c]{@{}c@{}}avg 5.76\%\\max 42.9\%\end{tabular} &
			0 &
			\begin{tabular}[c]{@{}c@{}}866.7K\\ 3231 models\\ 27.7\%\end{tabular} &
			0.76s \\ \hline
			\rowcolor[HTML]{FFFFFF} 
			Star &
			11942 &
			11567 &
			\begin{tabular}[c]{@{}c@{}}375\\ 86.1\% done\end{tabular} &
			308.1M &
			\begin{tabular}[c]{@{}c@{}}149.5M\\48.5\%\end{tabular} &
			\begin{tabular}[c]{@{}c@{}}158.6M\\51.5\%\end{tabular} &
			\begin{tabular}[c]{@{}c@{}}43.2M\\27.2\%\end{tabular} &
			\begin{tabular}[c]{@{}c@{}}901.9K\\5.69\%\end{tabular} &
			\begin{tabular}[c]{@{}c@{}}avg 18.1\%\\max 56.1\%\end{tabular} &
			0 &
			\begin{tabular}[c]{@{}c@{}}791.5K\\ 3780 models\\ 32.7\%\end{tabular} &
			0.83s \\ \hline
			\rowcolor[HTML]{EFEFEF} 
			TOT &
			35826 &
			34892 &
			\begin{tabular}[c]{@{}c@{}}935\\ 92.2\% done\end{tabular} &
			934.9M &
			\begin{tabular}[c]{@{}c@{}}453.8M\\48.5\%\end{tabular} &
			\begin{tabular}[c]{@{}c@{}}481.1M\\51.5\%\end{tabular} &
			\begin{tabular}[c]{@{}c@{}}74.0M\\15.4\%\end{tabular} &
			\begin{tabular}[c]{@{}c@{}}2.7M\\5.68\%\end{tabular} & 
			\begin{tabular}[c]{@{}c@{}}avg 9.5\%\\max 56.1\%\end{tabular} &
			0 &
			\begin{tabular}[c]{@{}c@{}}2.5M\\ 10177 models\\ 29.1\%\end{tabular} &
			0.78s \\ \hline
		\end{tabular}%
	}
	\vspace{1em}
	\caption{Summary of the results obtained in our large scale benchmark. In the second section of the table we report: the total number of input meshes; the number of maps successfully completed and the number of timeouts (i.e., the amount of runs that were stopped because an advancing move took more than $2s$). For timeouts, we also report the average percentage of triangles already inserted in the target domain when the process was stopped. In the second section of the table we report: the total number of advancing moves; the number and percentage of triangle flips and edge flips, and the number and percentage of convexifications and concavifications that were executed to unlock an illegal edge flip move. We then report the average and maximum growth rate of the meshes, measured as $(\vert T \vert_{out} - \vert T \vert_{in})/\vert T \vert_{in}$, and the number of flipped elements in the map, both with rationals and floating point coordinates. Flips in floating point coordinates were computed by forcing, at the end of the execution, a naive snap rounding to double for all mesh vertices. The total number of models containing at least one inverted element and the percentage w.r.t. the total number of converged models are also reported. Finally, in the rightmost column we report the average running time. Statistics on moves, growth, flips and run times do not consider timeouts. Including timeout experiments, more than one billion advancing moves were successfully performed by AFM.}
	\label{tab:benchmark}
\end{table*}

\section{Implementation Details}
\label{sec:implementative details}

The constructions described in the previous two sections theoretically guarantee both the existence and the injectivity of any mapping to a convex or star-shaped domain. The proposed approach entirely focuses on mesh \emph{validity} and does not make any attempt to generate an embedding with good geometric quality. As a result, output meshes can be expected to contain badly shaped and nearly degenerate triangles, which may be problematic to handle in a real software implementation. In this section we provide low level practical details about AFM, explaining how to transfer its correctness guarantees from the theory to a practical software tool.

\paragraph*{Numerical models.} Many nearly degenerate triangles will be created by AFM and, based on properties computed on such triangles, many algorithmic choices will be made during the execution. Expecting a correct algorithmic flow using limited precision floating points for calculation is practically impossible. 
Since all points inserted in the domains are computed by means of barycentric interpolation or by the computation of intersections, we know that all point coordinates will be rational numbers. We therefore implemented AFM using exact constructions to express both vertex coordinates and intermediate quantities, ensuring that no approximations will occur during the map computation. Specifically, we used the GMP~\cite{granlund2010gnu} rational numbers enhanced with the lazy evaluation scheme of CGAL~\cite{pion2011generic}.

\paragraph*{Snap rounding.} Despite enhanced by a lazy kernel, rational numbers are known to introduce dramatic slowdowns in the computation~\cite{CLSA20,cherchi2022interactive}. Even worse, due to our progressive approach vertex insertion is \emph{incremental}, thus introducing a cascading effect that may produce rational numbers with huge complexity. To mitigate the cascading problem and reduce the computational overhead introduced by our numerical model we always try to convert the coordinates of newly inserted vertices into floats. This is done by attempting a naive snap rounding, using \verb|CGAL::to_double()|, and then checking whether the rounded coordinates have inverted the orientation of any triangle incident to such vertex. If no flip is found, the snap rounding is accepted. Otherwise, the coordinates for that specific vertex are kept rational. Based on our experience, this simple strategy reduces running times by almost one order of magnitude.


\paragraph*{Algorithmic choices.} AFM has been explicitly designed to operate with rational numbers. This is perhaps visible in the geometric constructions that we used for convexification (\cref{sssec:convexification}) and concavification (\cref{sssec:concavification}), which entirely avoid the use of normalized vectors (which involve the computation of squared roots) and trigonometric functions. Line searches for vertex repositioning were also avoided because when implemented in infinite precision may produce (almost) infinite loops even in cases when convergence is guaranteed from a theoretical perspective.
Alternative geometric constructions may have possibly produced better shaped triangles or smaller vertex displacements, but our approach entirely based on the computation of segment intersections is fully compatible with our numerical model and is \emph{closed form}, in the sense that it is based on finite quantities that can be readily computed, without requiring iterations or numerical algorithms for their estimation.

\section{Results}
\label{sec:results}
\begin{figure*}
	\includegraphics[width=\linewidth]{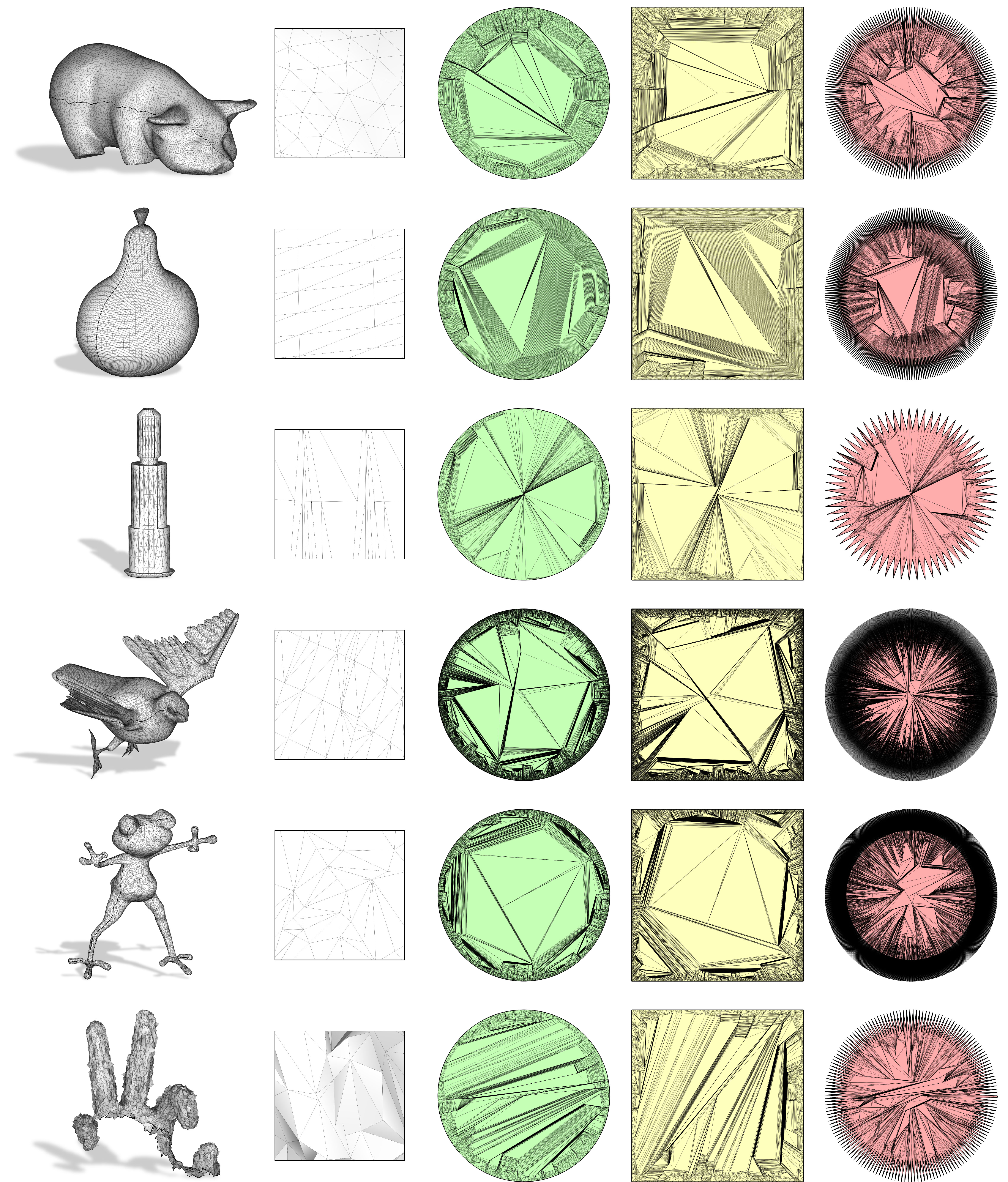}
	\caption{Results obtained with AFM on a variety of meshes with different geometric and topological properties. All maps are fully injective.}
	\label{fig:mosaic}
\end{figure*}

We implemented AFM in C++, using Cinolib~\cite{livesu2019cinolib} for geometry processing and \verb|CGAL::Lazy_exact_nt<CGAL::Gmpq>| for exact computations. Our software prototype is currently implemented as a single threaded application, supporting mappings to circles, squares and star-like polygon domains as the ones shown in~\cref{fig:teaser,fig:mosaic}. Our reference implementation will be released to the public domain upon acceptance of the article. Code improvements, such as parallelization and warm starting, are possible and would likely greatly boost our performances, but are not currently implemented. Considerations about these future extensions can be found in~\cref{sec:conclusions}.

\paragraph*{Dataset}
To validate our approach we conducted a large-scale experimentation involving roughly 12K meshes collected from the 2D Structure Dataset~\cite{carlier20162d} and from data released by the authors of~\cite{liu2018progressive,chai2018sphere}. We composed this testing dataset so as to be heterogeneous in multiple ways, exposing AFM to a wide variety of mesh sizes (from 190 to 230K triangles), shapes (humans, animals, abstract objects, tools, furnitures, silhouettes), surface types (smooth, noisy) and local mesh structures (regular, irregular, high vertex valence). 

\paragraph*{Setup} Overall, we run 36K experiments, launching AFM three times for each input model, producing mappings to three alternative target domains, both convex and star-shaped. All tests have been executed on commodity hardware (a Mac Book M1 Pro with 32GB of RAM). Considering the limited hardware capabilities at our disposal, we decided to set a strict time limit for each experiment, aborting the map generation if a single advancing move took more than 2 seconds. Comprehensive statistical information for all our tests are reported in \cref{tab:benchmark}.

\paragraph*{Baseline}
We also considered alternative provably robust methods for comparative analysis. Our baseline is composed by Tutte~\shortcite{tutte1963draw} and Progressive Embedding (PE)~\cite{shen2019progressive}. For the former, we considered the implementation available in Cinolib~\cite{livesu2019cinolib}. For the latter, we used the reference implementation released by the authors. Both implementations use floating point numbers and PE also exploits parallelization to speed up computation. The reference code for PE assumes to receive in input a previously existing mapping, which is only locally modified to remove inverted elements (if any). The authors released two tools to warm start PE: one based on Tutte and one based on randomization. Since in most of the cases Tutte already computes an injective map, bootstrapping PE with Tutte would trigger the execution of PE on a tiny amount of cases (5 out of 12K tests). We therefore initialized an invalid embedding using their \verb|random_init_bin| tool and then processed the so generated file with their tool \verb|untangle_bin|, using option \verb|-e 1|. While better initial solutions would certainly result in higher performances, this is the only setup that allowed use to ensure that PE is indeed executed on every single input mesh. 

We tested the baseline only for mappings to a strictly convex domain (circle). Star-like domains have been discarded because these methods are less general than AFM and do not support concavities. Square domains have also been discarded, because these methods are not able to change the input mesh connectivity to open the space of solutions. Therefore, they may be asked to solve an impossible problem if the input mesh contains triangles whose vertices are all on the boundary and map to the same side of the square, becoming co-linear.

Due to the excessive computational cost of PE and the limits of our testing hardware, we could not complete all tests for this method. We stopped the execution after 24h of computation. At that point, only 139 out of the 12K maps were completed. The evaluation of PE is therefore limited to this restricted set of shapes. Conversely, Tutte was executed on the full set of 12K input shapes, which completed in 8.3 minutes.

\begin{figure}
	\includegraphics[width=\columnwidth]{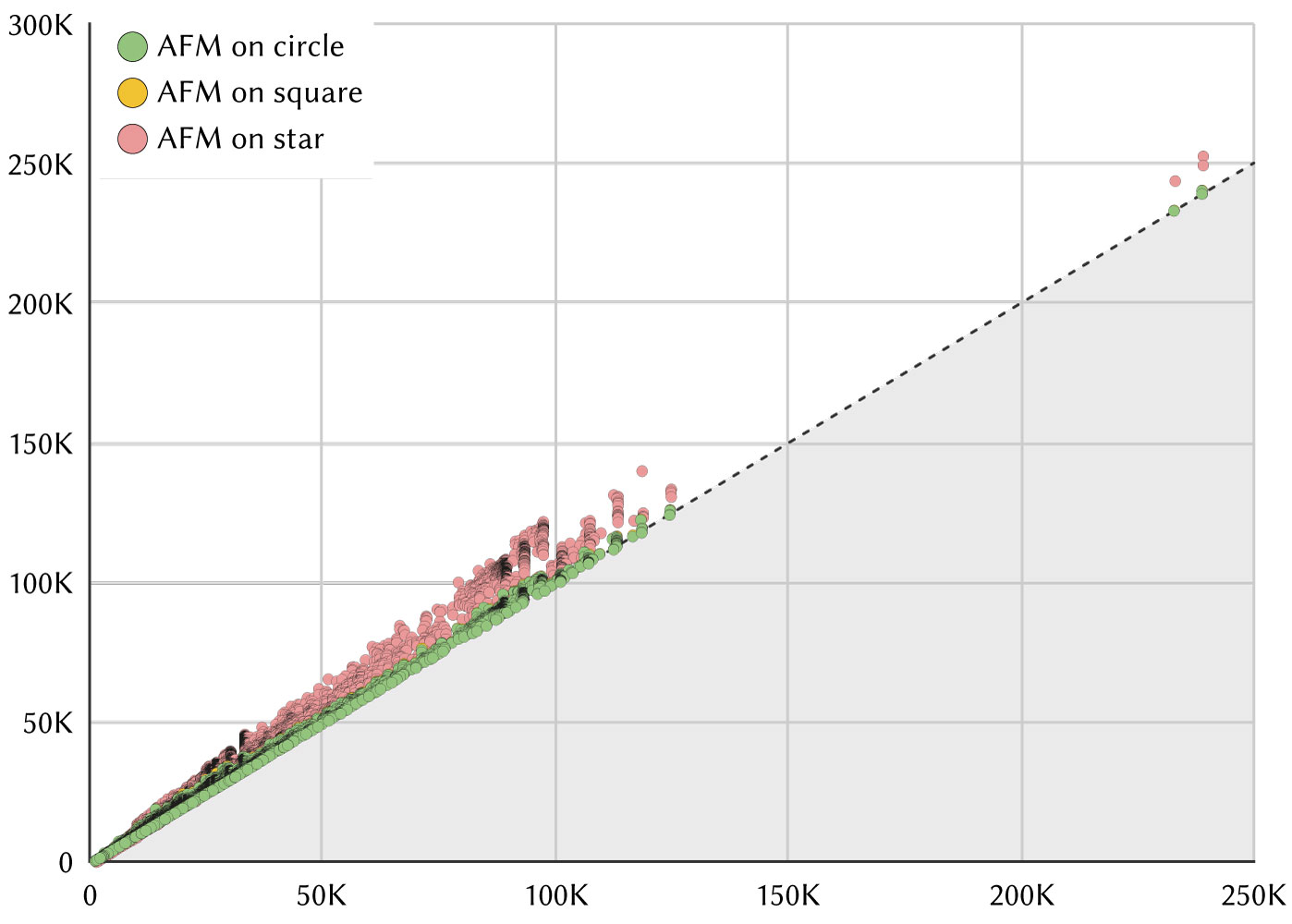}
	\caption{Plot of output mesh size (vertical) over input mesh size (horizontal) for our 36K mapping tests. Points on the dashed line correspond to maps that introduced zero refinement. AFM only marginally refines the input mesh: mappings to circles and squares perfectly overlap and are almost entirely flat on the dashed line. Mappings to concave (star-shaped) domains are more likely to trigger refinement, but remain just slightly above it.}
	\label{fig:growth}
\end{figure}

\begin{figure}
	\includegraphics[width=\columnwidth]{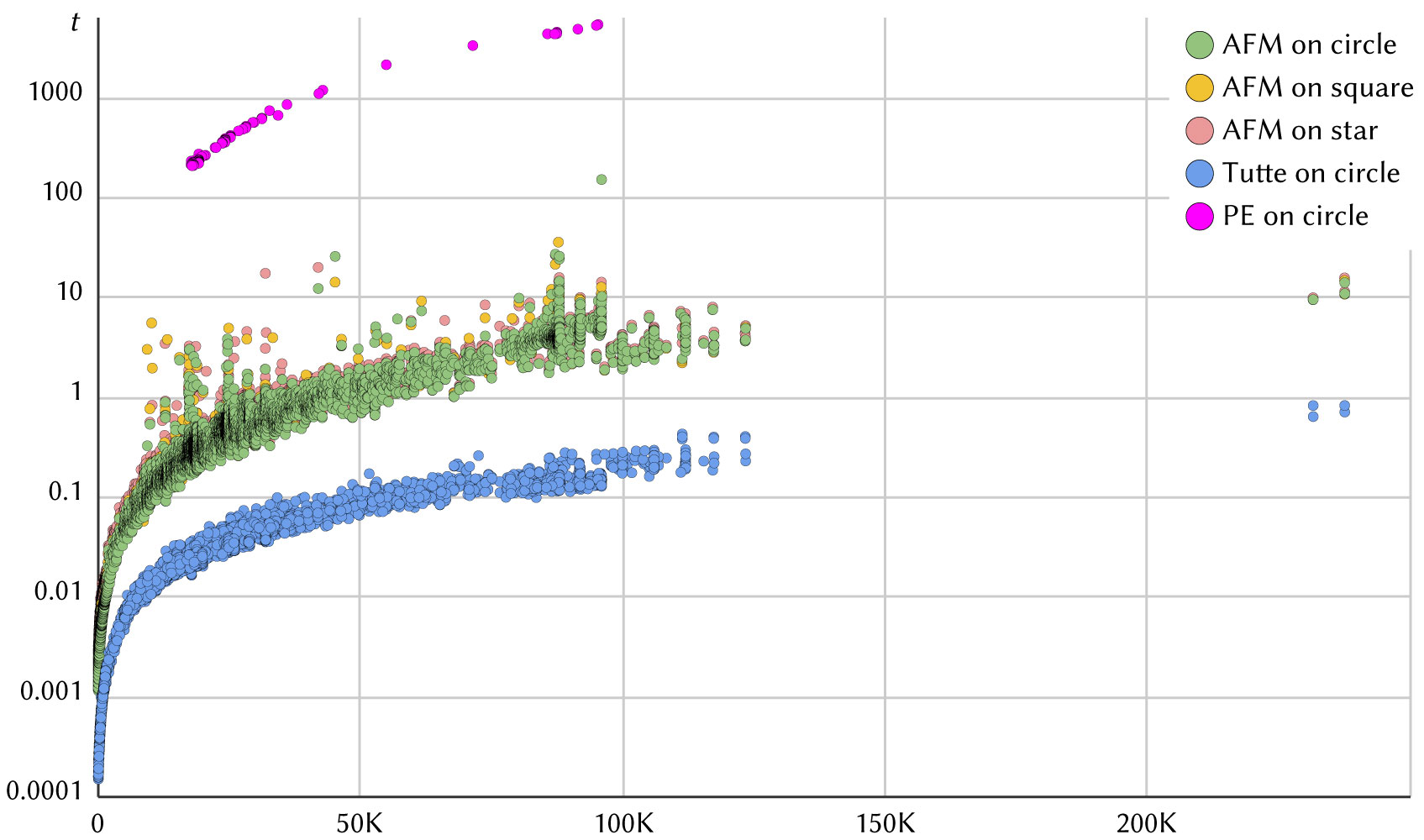}
	\caption{Running times (in seconds) for AFM, Tutte~\shortcite{tutte1963draw} and Progressive Embedding~\cite{shen2019progressive} on our testing dataset. For AFM we considered mappings to circles, squares and stars. Since Tutte and PE do not support adaptive refinement we only tested mappings to a strictly convex domain. Due to the prohibitive running time of PE on our commodity laptop, tests for this method were stopped after 24h of computation, when only 139 maps were completed. On average, Tutte is one order of magnitude faster than our method, which in turn is three orders of magnitude faster than PE.}
	\label{fig:timings}
\end{figure}

\paragraph*{Timings} Despite the single thread implementation and the use of computationally expensive rational numbers, on average AFM was able to complete each mapping task in less than a second (avg $0.78s$). As shown in \cref{fig:timings}, floating point Tutte is by far the fastest algorithm (avg $0.04s$), one order of magnitude faster than AFM. PE is much slower (avg $608s$), three and four orders of magnitude slower than AFM and Tutte, respectively.
 
For AFM, running times remain consistent across the target domains we considered, with mappings to circles and squares being slightly faster (avg $0.76s$) than mappings to stars (avg $0.83s$). This difference can be explained by observing that star mappings tend to refine more (\cref{fig:growth}), thus introducing a computational overhead. 

In a few cases our tool was stopped because the insertion of a triangle took more than two seconds, violating our testing policy. As can be noticed in~\cref{tab:benchmark} (percentages in the Timeout column), stopped processes had already completed more than the $92\%$ of the map on average, meaning that only a handful of input triangles were missing in the target domain.
Based on our analysis, this slowdown towards the end of the execution is entirely due to cascading issues with the rational numbers, which accumulate complexity throughout the computation. The snap rounding strategy discussed in \cref{sec:implementative details} is naive and it only mitigates this issue, without entirely avoiding it.
Possible remedies are discussed in \cref{sec:conclusions}.

\paragraph*{Robustness}
Thanks to our rational implementation AFM is fully robust. In the 36K runs it was tested on, more than one billion advancing moves were successfully completed without ever introducing a degenerate or inverted triangle in the target domain, thus confirming the correctness of our topological and geometric constructions and their practical robustness. As detailed in \cref{tab:benchmark}, this quantity is almost perfectly split into triangle splits and edge flips, with just a slight predominance of the latter over the former.

Another important aspect to consider when evaluating the practical robustness of a mapping method is its ability to not get stuck due to a vanishing space of solutions. A key advantage of AFM over prior provably robust methods is its ability to exploit local mesh refinement to create a valid solution even when it does not exist (\cref{fig:star}). While basic topological checks like the 2-connectedness of the graph could be executed in a pre-processing phase and remove some of the potential issues (\cref{sssec:refinement}), mappings to concave domains may require non trivial refinement of the input mesh that is hard to determine a priori. This is also confirmed by our statistical results, which indicate that AFM used more refinement to complete mappings to star domains than the one necessary to map to circles and squares. Specifically, mesh growth for star maps was  $18.1\%$ on average, whereas growth for circles and squares was $4.91\%$ and $5.76\%$, respectively. In the worst case the amount of refinement reached a peak of $56.1\%$ of the input mesh size. Nevertheless, our analysis revealed that the highest percentages of mesh growth belong to very coarse meshes containing either none of just a few internal vertices and edges, such as the one shown in \cref{fig:star}. This is confirmed by the full plot in \cref{fig:growth}, which shows that for the 24K maps to circles and squares input and output mesh sizes were almost identical, wheres for the 12K maps to stars output mesh size was only marginally higher than the input one.

%
%
%
%
%

\begin{table}[]
	\resizebox{\linewidth}{!}{%
\begin{tabular}{r|c|c|c|c|c|}
	&
	\begin{tabular}[c]{@{}c@{}}Guaranteed\\ Injective\end{tabular} &
	\begin{tabular}[c]{@{}c@{}}Floating Point\\ Robust\end{tabular} &
	\begin{tabular}[c]{@{}c@{}}Running\\ Time\end{tabular} &
	\begin{tabular}[c]{@{}c@{}}Supported\\ Domains\end{tabular} &
	\begin{tabular}[c]{@{}c@{}}Smart\\ Refinement\end{tabular} \\ \hline
	Tutte &
	\cellcolor[HTML]{67FD9A}{\color[HTML]{333333} yes} &
	\cellcolor[HTML]{FFCC67}{\color[HTML]{333333} \begin{tabular}[c]{@{}c@{}}99.96\%\\ \footnotesize{11946/11950}\end{tabular}} &
	\cellcolor[HTML]{67FD9A}{\color[HTML]{333333} \begin{tabular}[c]{@{}c@{}}0.04s\\ \footnotesize{serial, float}\end{tabular}} &
	\cellcolor[HTML]{FFCC67}{\color[HTML]{333333} \begin{tabular}[c]{@{}c@{}}convex\\ \footnotesize{only if a map exists}\end{tabular}} &
	\cellcolor[HTML]{FD6864}{\color[HTML]{FFFFFF} no} \\ \hline
	PE &
	\cellcolor[HTML]{67FD9A}{\color[HTML]{333333} yes} &
	\cellcolor[HTML]{67FD9A}{\color[HTML]{333333}  \begin{tabular}[c]{@{}c@{}}100\%\\ \footnotesize{139/139}\end{tabular}} &
	\cellcolor[HTML]{FD6864}{\color[HTML]{FFFFFF} \begin{tabular}[c]{@{}c@{}}608s\\ \footnotesize{parallel, float}\end{tabular}} &
	\cellcolor[HTML]{FFCC67}{\color[HTML]{333333} \begin{tabular}[c]{@{}c@{}}convex\\ \footnotesize{only if a map exists}\end{tabular}} &
	\cellcolor[HTML]{FD6864}{\color[HTML]{FFFFFF} no} \\ \hline
	AFM &
	\cellcolor[HTML]{67FD9A}{\color[HTML]{333333} yes} &
	\cellcolor[HTML]{FD6864}{\color[HTML]{FFFFFF} \begin{tabular}[c]{@{}c@{}}70.8\%\\ \footnotesize{24715/34892}\end{tabular}} &
	\cellcolor[HTML]{FFCC67}{\color[HTML]{333333} \begin{tabular}[c]{@{}c@{}}0.78s\\ \footnotesize{serial, rational}\end{tabular}} &
	\cellcolor[HTML]{67FD9A}{\color[HTML]{333333} star-shaped} &
	\cellcolor[HTML]{67FD9A}{\color[HTML]{333333} yes} \\ \hline
\end{tabular}%
	}
	\vspace{1em}
	\caption{Summary of the outcomes of our large-scale comparative analysis. Percentages in the second column refer to the amount of fully injective mappings in double precision produced by each method over the totality of the attempts. Third column reports average running times on the same attempts. Note that due to prohibitive running times PE was tested on only 139 models, produced in 24h. Tutte was executed 12K times (mapping to a circle) and AMF was executed 36K times (mapping to circles, squares and stars, 12K times each). For the supported domains column, Tutte and PE permit maps to non strictly convex domains (e.g. a square) only if the mesh does not contain triangles having all three vertices on the boundary. Conversely, AFM exploits (smart) local mesh refinement to automatically open the space of solutions in case a mapping does not exist.}
	\label{tab:summary}
\end{table}
\section{Conclusions and Future Works}
\label{sec:conclusions}
We have introduced Advancing Front Mapping (AFM), a novel constructive method for the provably robust computation of injective mappings to convex and star-shaped domains. AFM expands the (rather small) set of unconditionally robust tools for surface mapping, also providing novel capabilities which are not found in prior similar methods. The two key theoretical advancements that AFM brings into this field are: the possibility to robustly compute maps to star-shaped domains, and the ability to exploit local mesh refinement to create a valid solution in case the connectivity of the input mesh does not permit to create one. The robustness, performances, and ability to our method to sustain all these claims have been extensively validated in a large-scale experimentation, testing our software implementation on 36K input cases. 

The results of our comparative analysis, conveniently summarized in \cref{tab:summary}, emphasize that in addition to these positive aspects, AFM also falls behind prior art in terms of efficiency, where the Tutte~\shortcite{tutte1963draw} embedding remains unbeaten and is one order of magnitude faster, and in terms of ability to produce a valid floating point map, a task where Progressive Embedding~\cite{shen2019progressive} excels.

It should be noted that none of these features are the core business of this article, which is rather focused on the presentation of this novel construction and its empirical validation. In the remainder of this section we share a few considerations on future extensions of our algorithm that may provide significant benefits in terms of these and other aspects, both in 2D and 3D. 

\paragraph{Parallelization} Our current implementation of AFM relies on a serial mesh data structure, hence is single threaded. From an algorithmic perspective AFM could be trivially parallelized by applying simultaneous advancing moves in different regions of the active front. Designing fully disjoint advance operations is easy, because each move has a very local footprint. If the move is a triangle split, the only mesh element that changes is the triangle being split. If the advancing move is an edge flip, then two adjacent front edges plus the already inserted triangles that are incident to any of the three front vertices involved may change (due to the possibility of local refinement). Considering the 10 cores of our testing hardware, using a parallel-friendly mesh data structure such as~\cite{jiang2022declarative} we could reasonably expected a boost factor of 4-5$\times$.

\paragraph{Warm Start} AFM is currently designed to create a surface map from scratch, always inserting \emph{all} the input triangles in the target domain. Alternative methods such as PE hugely benefit from a warm start initialization (e.g., computed with Tutte), which permits to focus only on the very few spots where the map needs to be fixed. AFM could be similarly modified to enjoy a warm start, identifying the inverted triangles in the input map and then defining small star-shaped neighborhoods that fully contain each of them (e.g., via flooding). At this point, each of these regions could become a separated active front, and the triangles inside each region being re-inserted with our advancing front methodology. In the worst case scenario, the local neighborhood could grow to conquer the whole mesh, yielding a problem identical to the one we currently solve. Conversely, if smaller sized star-shaped pockets around each illegal triangle are successfully found, each of them could be processed separately, reducing the computational cost and also providing yet another opportunity for parallelization.

\paragraph{Indirect Predicates and Cascading} As shown in multiple recent articles~\cite{cherchi2022interactive,CLSA20}, switching from rational numbers to Indirect Predicates~\cite{attene2020indirect} provides great advantages in terms of efficiency without sacrificing robustness. Such a change should provide similar advantages also in our setting. Unfortunately, Indirect Predicates do not support cascading, hence the switch is currently incompatible with our implementation. A tempting idea to resolve the cascading problem consists in expressing all mapped points as a convex combination of the vertices of the polar mesh computed during initialization (\cref{sssec:background_mesh}). Specifically, since such a mesh is a valid tessellation of the target domain, any mesh point is strictly contained either in one of its triangles or edges. Devising alternative construction that allow to exactly represent split point locations with only one indirection is an appealing direction of further research.

\paragraph{Snap Rounding and Regularization} In their current form, the mappings generated with AFM cannot be passed to a downstream application because vertex coordinates are rational and not floating point. Switching from one numerical model to the other in a robust manner is a notoriously complex problem, called 3D snap rounding~\cite{devillers20183d}. Besides the naive rounding to double described in \cref{sec:implementative details}, our method does not make any serious attempt to robustly generate floating point maps, failing at this task in the $29.2\%$ of the cases (\cref{tab:summary}). Provably correct snap rounding algorithms are still too complex to be practically useful. Effective heuristics often used in the context of mesh arrangements (\cite{zhou2016mesh} \S6.1) are not usable in our context because they require to collapse triangles in the input mesh, failing to preserve its original geometry. The only possible solution that we envision is to incorporate in our pipeline a regularization step that relaxes the geometry, avoiding almost degenerate configurations. This routine has been successfully used by Progressive Embedding~\cite{shen2019progressive}, although in their case it introduces a huge computational overhead because vertex insertions may arise \emph{anywhere} in the mesh, requiring a global optimization. Our approach operates only along the active front, keeping the geometry of the previously inserted triangles frozen. Introducing vertex relaxation only along the current front may permit to find a sweet spot between floating point robustness and numerical overhead. Experiments in this direction have not been attempted yet.


\begin{figure}
\includegraphics[width=\columnwidth]{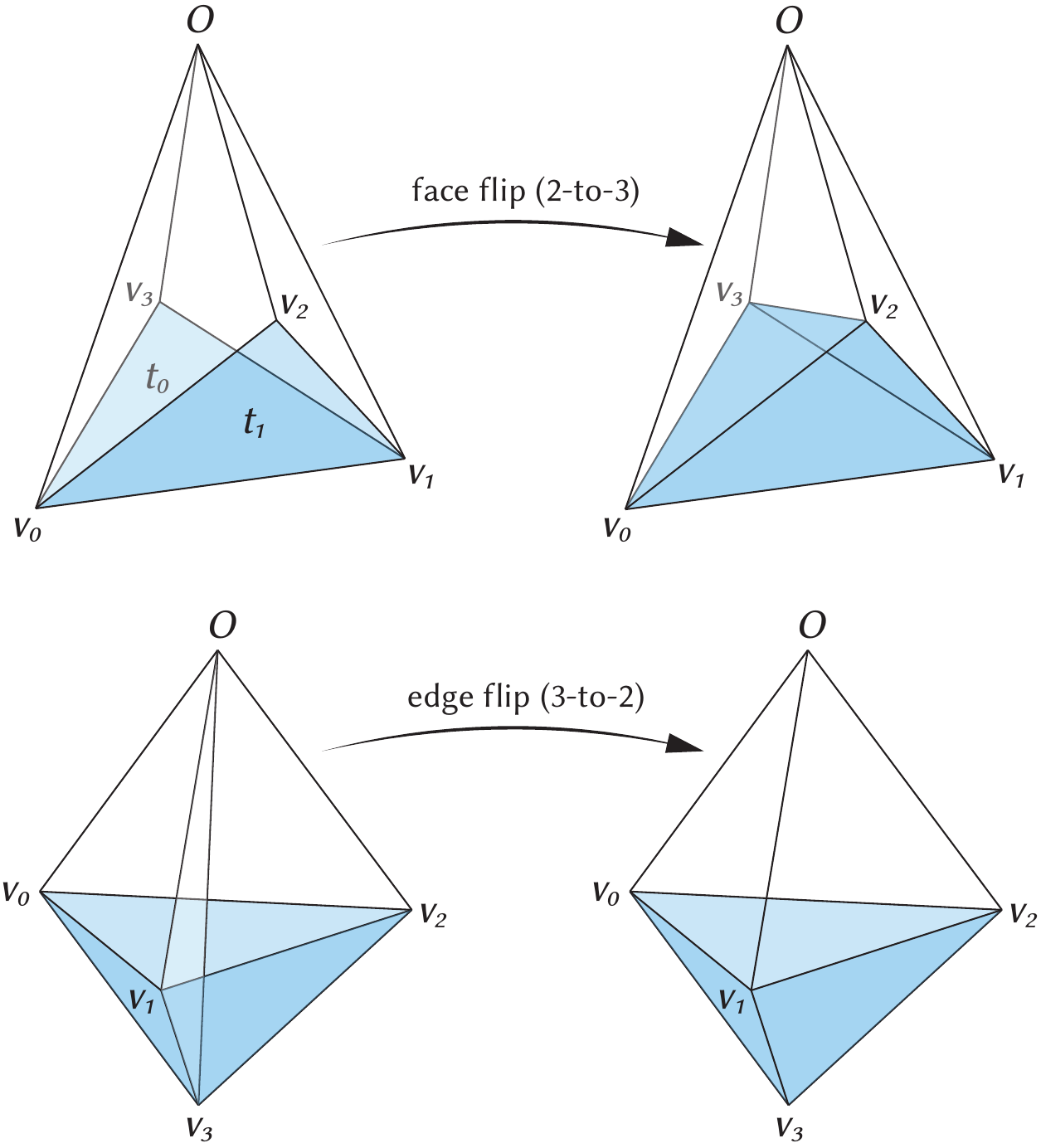}
\caption{Advancing front moves in a volume mesh by means of flipping operators. Top: tetrahedra having two triangular faces exposed on the front ($t_1,t_2$), can be inserted in the domain by flipping the triangle $v_0,v_1,O$. Bottom: tetrahedra having three faces exposed on the front can be reproduced by flipping edge $v_3,O$.}
\label{fig:tetflips}
\end{figure}

\paragraph{Volume Maps} Last but not least, an appealing property of AFM is its apparent compatibility with a volumetric extension. Provably injective volume mapping is a fundamental yet open problem in the literature, for which existing robust 2D approaches are either known to have severe restrictions~\cite{alexa2023tutte} or are unlikely to extend (\cite{livesu2020mapping} \S2). Regarding AFM, the generation of the polar mesh in the initialization phase (\cref{sec:init}) and the basic advancing moves by means of triangle splits and edge flips (\cref{sec:advancing_moves}) have a direct counterpart in tetrahedral meshes, with the only difference that in 3D there are two alternative flip operations to advance the front: a face flip move to conquer a tetrahedron having two faces on the front and an edge flip move to conquer a tetrahedron having three faces on the front (\cref{fig:tetflips}). What remains unclear -- and is subject to ongoing research -- is the handling of locally concave configurations that prevent the execution of a flip. As discussed in \cref{sec:corner_cases} in 2D there are only two possible cases, that are fully addressed by our convexification and concavification strategies, whereas in 3D there are more, possibly more complex.

\begin{acks}
This paper is the result of a work that lasted almost five years. During this long period of time I have shared my ideas and thoughts with many friends and colleagues. Thanks are due to Daniele Panozzo, Teseo Schneider and Zhongshi Jiang, with whom I failed to make a preliminar version of this idea work back in 2019. I am particularly grateful to Gianmarco Cherchi, Enrico Puppo, Marco Attene and Riccardo Scateni, who have patiently listened to me spelling the steps of this algorithm for years, often pointing me in the right direction to solve unexpected corner cases.
\end{acks}

\bibliographystyle{ACM-Reference-Format}
\bibliography{biblio}



\end{document}